# Molecular Dynamics Simulation of Smaller Granular Particles Deposition on a Larger One Due to Velocity Sequence Dependent Electrical Charge Distribution

[a]Euis Sustini, [b]Siti Nurul Khotimah, [a]Ferry Iskandar, and [b,*]Sparisoma Viridi

[a]Physics of Electronic Materials Research Division
Institut Teknologi Bandung, Jalan Ganesha 10, Bandung 40132, Indonesia
[b]Nuclear Physics and Biophysics Research Division
Institut Teknologi Bandung, Jalan Ganesha 10, Bandung 40132, Indonesia
[*]Email: dudung@fi.itb.ac.id

Abstract

Deposition of smaller granular particles on a larger nucleus particle has been simulated in two-dimension using molecular dynamics method. Variation of sequences of velocity of deposited particles is conducted and reported in this work. The sequences obey a normal distribution function of velocity with the same parameters. It has been observed that for velocity in range of 0 to 0.02 the densest deposited site (15-17 % number of grains) is located at about angle $\pi/4$ where location of injection point is $\pi/4$. And the less dense is about $\pi/4 + \pi/2$. Different sequences give similar result.

## Introduction

The mechanism of how smaller particles attached to a large particle containing different surface charge is still an interesting thing to investigate [1-5]. In the powder processing, including their fabrications and modifications, the interaction of inter-particles becomes very important to optimize the product. The processes are including the nanoparticle coating and controlling agglomeration that is the most important thing among the powder processes. Since the particles that used in the process are typically has electrical charge, therefore, to understanding how smaller particle attached in the larger particles due to electrical force is very important. Here we report on the using of molecular dynamics method that implemented Gear predictor-corrector algorithm [6] to simulate the deposition of smaller particles on a larger particle. We believe that this study becomes useful to understand the mechanism of particle agglomeration in aerosol process as well as the mechanism of mixture or coating process in the functionalization of powder. In addition, it also will be useful in evaluation of charge distribution on the surface of larger particle using the deposition of smaller particles, for example, in evaluation of the charge distribution of toner powder.

## Simulation procedure

Two types of force are considered in this simulation, which are normal force and electrostatic force. The former type of force acts as repulsive force in order to hinder two grains collapse into each other and the later acts as repulsive and attractive force between two charged grains. The normal force has formulation which is known as linear dash-pot model [7]

$$\vec{N}_{ij} = k_r \xi_{ij} \hat{r}_{ij} - k_v \dot{\xi}_{ij} \hat{r}_{ij}, \quad (1)$$

where $\xi_{ij}$ is defined as overlap between two grains and $\dot{\xi}_{ij}$ is the derivative of $\xi_{ij}$ with respect to time $t$. The overlap $\xi_{ij}$ is

$$\xi_{ij} = \max\left[0, \frac{1}{2}(d_i + d_j) - r_{ij}\right]. \quad (2)$$

And the electrostatic force has formulation

$$\vec{Q}_{ij} = \hat{r}_{ij}\left(k_q \frac{q_i q_j}{r_{ij}^2}\right), \quad (3)$$

Both Equations (1) and (3) used following definitions

$$\vec{r}_{ij} = \vec{r}_i - \vec{r}_j, \quad (4)$$

$$r_{ij} = |\vec{r}_{ij}| = \sqrt{\vec{r}_{ij} \cdot \vec{r}_{ij}}, \quad (5)$$

$$\hat{r}_{ij} = \frac{\vec{r}_{ij}}{r_{ij}}. \quad (6)$$

Gear predictor-corrector algorithm of fifth order [6] is chosen in the molecular dynamics method used in the simulation, which has two steps: prediction step (written with upper index $p$) and correction step for every particular grain. The first step is formulated as

$$\begin{pmatrix} \vec{r}_0^{\,p}(t+\Delta t) \\ \vec{r}_1^{\,p}(t+\Delta t) \\ \vec{r}_2^{\,p}(t+\Delta t) \\ \vec{r}_3^{\,p}(t+\Delta t) \\ \vec{r}_4^{\,p}(t+\Delta t) \\ \vec{r}_5^{\,p}(t+\Delta t) \end{pmatrix} = \begin{pmatrix} 1 & 1 & 1 & 1 & 1 & 1 \\ 0 & 1 & 2 & 3 & 4 & 5 \\ 0 & 0 & 1 & 3 & 6 & 10 \\ 0 & 0 & 0 & 1 & 4 & 10 \\ 0 & 0 & 0 & 0 & 1 & 5 \\ 0 & 0 & 0 & 0 & 0 & 1 \end{pmatrix} \begin{pmatrix} \vec{r}_0(t) \\ \vec{r}_1(t) \\ \vec{r}_2(t) \\ \vec{r}_3(t) \\ \vec{r}_4(t) \\ \vec{r}_5(t) \end{pmatrix}. \quad (7)$$

And the correction step will give the corrected value of $\vec{r}_n(t+\Delta t)$ through

$$\begin{pmatrix}\vec{r}_0(t+\Delta t)\\ \vec{r}_1(t+\Delta t)\\ \vec{r}_2(t+\Delta t)\\ \vec{r}_3(t+\Delta t)\\ \vec{r}_4(t+\Delta t)\\ \vec{r}_5(t+\Delta t)\end{pmatrix}=\begin{pmatrix}\vec{r}_0^{\,p}(t+\Delta t)\\ \vec{r}_1^{\,p}(t+\Delta t)\\ \vec{r}_2^{\,p}(t+\Delta t)\\ \vec{r}_3^{\,p}(t+\Delta t)\\ \vec{r}_4^{\,p}(t+\Delta t)\\ \vec{r}_5^{\,p}(t+\Delta t)\end{pmatrix}+\begin{pmatrix}c_0\\ c_1\\ c_2\\ c_3\\ c_4\\ c_5\end{pmatrix}\Delta\vec{r}_2(t+\Delta t),\quad(8)$$

with

$$\Delta\vec{r}_2(t+\Delta t)=\vec{r}_2(t+\Delta t)-\vec{r}_2^{\,p}(t+\Delta t).\quad(9)$$

The term $\vec{r}_n(t+\Delta t)$ is defined as

$$\vec{r}_n(t)=\frac{(\Delta t)^n}{n!}\left[\frac{d^n\vec{r}_0(t)}{dt^n}\right],\quad(10)$$

where $\vec{r}_0$ is position of a grain. The term $\vec{r}_2(t+\Delta t)$ in correction term in Equation (9) is obtained from Newton's second law of motion. For example, particle $i$ has

$$\begin{aligned}\left[\vec{r}_2(t+\Delta t)\right]_i &= \frac{(\Delta t)^2}{m_i}\\ &\times\sum_{j\neq i}\vec{Q}_{ij}(t+\Delta t)+\vec{N}_{ij}(t+\Delta t)\end{aligned}.\quad(11)$$

The left part of Equation (11) is calculated using $\vec{r}_n^{\,p}(t+\Delta t)$.

**Model of smaller grains deposition**

A larger grain (nucleus grain) is positioned at a fixed position and smaller grains (deposited grains) have identical mass, size, and charge. The negatively charged smaller grains, in certain sequence of velocity, are injected to the system at fixed entry point as illustrated in Figure 1.

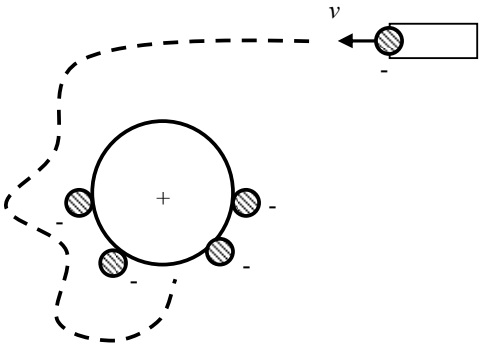

Figure 1. Smaller grains are injected to the system at a fixed point, attracted to larger grain by electrostatic force, and deposited to larger grain at some positions.

Through variation of grain sequence in velocity, sites of deposition will be different even for same normal distribution of velocity.

A statistical calculation will accompany simulation result in order to predict the probability of variation of deposition sites.

**Distribution and sequences**

A normal velocity distribution of deposited grains is chosen as starting points in the simulation. This normal distribution can written as

$$f(v)=\frac{1}{\sigma\sqrt{2\pi}}\exp\left[-\frac{(v-\bar{v})^2}{2\sigma^2}\right],\quad(12)$$

where $\bar{v}$ and $\sigma$ are average value of velocity and its standard deviation, respectively. This distribution satisfies that $\int_v f(v)dv=1$. The detail of how to produce discrete normal distribution function and its related sequences from Equation (12) can be found in [8] and the codes are also available [9].

For $\bar{v}=0.5$, $\sigma=0.15$, $\Delta v=0.1$, and $N=100$ an illustration of $\Delta N_v=Nf(v)\Delta v$ is given in Figure 2, where value of the function is drawn in solid line and its integer value is drawn using circle. For generating sequence of particle, number of particle is obtain from integration of Equation (12), but for simplicity we use only discreet value of $v$ as shown in Table 1, which gives number of particle about 100. And for certain value of $v$, number of particle can found in the most left column of Table 1. In this simulation only ten different value of $v$ will be used.

Table 1. Value of $v$, $f(v)$, and the integer value of $\Delta N_v=Nf(v)\Delta v$.

| $v$ | $f(v)$ | Int[$\Delta N_v$] |
|---|---|---|
| 0.0 | 0.0103 | 0 |
| 0.1 | 0.0760 | 1 |
| 0.2 | 0.3599 | 4 |
| 0.3 | 1.0934 | 11 |
| 0.4 | 2.1297 | 21 |
| 0.5 | 2.6596 | 26 |
| 0.6 | 2.1297 | 21 |
| 0.7 | 1.0934 | 11 |
| 0.8 | 0.3599 | 4 |
| 0.9 | 0.0760 | 1 |
| 1.0 | 0.0103 | 0 |
| $\bar{v}$ | 0.5 | |
| $\sigma$ | 0.15 | |
| $A$ | 2.65962 | |
| $N$ | 99.981 | 100 |

At $v=0.5$ value of $\Delta N$ is rounded down to achieve number of particles to be 100 instead of 101. For other values the common integer rounding rule are used.

The sequences are then produced using velocity values from Table 1. An initial sequence is simple as

$$\begin{array}{l}0.1,\ 0.2,\ 0.2,\ 0.2,\ 0.2,\ 0.3,\ ..,\ 0.4,\\ ..,\ 0.5,\ ..,\ 0.6,\ ..,\ 0.7,\ ..,\\ 0.8,\ 0.8,\ 0.8,\ 0.8,\ 0.9\end{array}\quad(13)$$

as explained in [8]. Using a flat random swap process the initial sequence in Equation (13) is changed to other sequence using a certain random seed [9]. Since the sequences are identified and labeled using the seed number, they are reproducible.

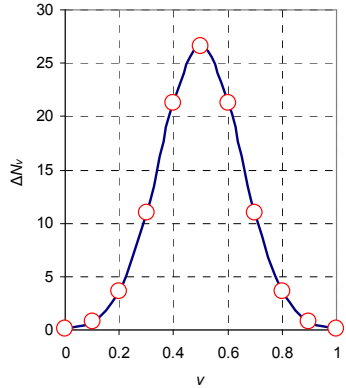

Figure 2. Illustration of discreet normal distribution of $v$ for $\bar{v} = 0.5$, $\sigma = 0.15$, which gives $A = 2.65962$.

Four examples of sequences that obey distribution function in Figure 2, are given in Figure 3. These sequences are produced from the initial sequence in Equation (13) using random seed 1, 2, 3, and 4.

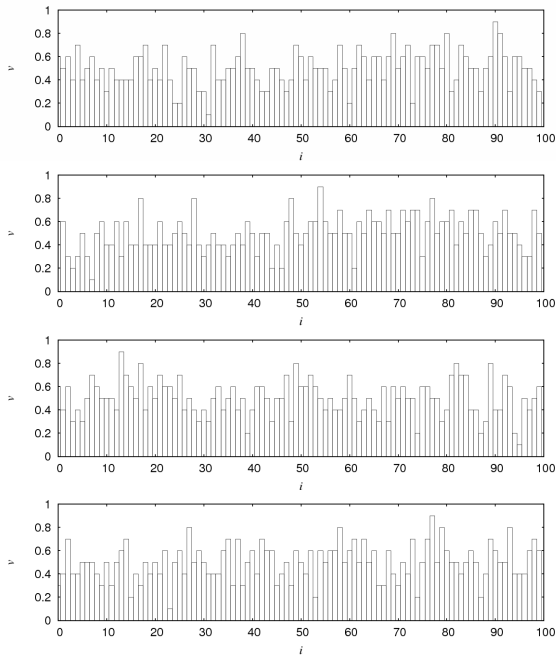

Figure 3. Four examples of produced random sequence of velocity $v$, with random seed 1, 2, 3, 4 for the sequence pictures from top to bottom.

The variable $i$ in Figure 3 indicates the order in the sequence. The mentioned flat random swap process is actually swapping this variable in producing other sequences from the initial sequence.

**Results and discussion**

A typical result of a deposited grain (written with index 1) and a nucleus grain (written with index 2) is given in Figure 4. Grain 1 is departed from (2, 2) with initial velocity (-0.01, 0). Its motion is caused by attractive electric force between grain 1 and grain 2.

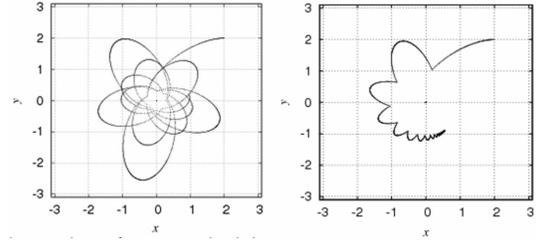

Figure 4. Trajectory of 0.1 m diameter deposited grain on 0.5 diameter nucleus positioned at (0, 0) grain with initial velocity of deposited grain (-0.01, 0) departed from (2, 2) for $k_v = 10^{-1}$ (left) and $k_v = 2$ (right).

A parameter search is done subjectively to get an optimum time to make the deposited grain stop bouncing. Figure 4 shows the result for $k_v = 10^{-1}$ and $k_v = 2$. We choose the latest result since it still gives more possibilities for the smaller grains to attach at different places instead of direct attachment, but it has smaller time to be attached than the first result. Then, following parameters are used in rest of the report: $k_q = 10^{-3}$, $k_r = 10^4$, $k_v = 2$, $t_i = 0$, $t_f = 2 \times 10^3$, $\Delta t = 10^{-3}$, $T_s = 10^{-1}$, $m_1 = 1$, $m_2 = 0.1$, $q_1 = 10^2$, $q_2 = -10^{-2}$, $d_1 = 2$, $d_2 = 0.5$ and $v' = 0.02v$, with $v'$ is the real physical velocity where $v$ is the velocity in the distribution function $f(v)$.

There is also an assumption that every time injecting a new small grain, previously attached grains are considered fixed at nucleus grain by sort of binding force. It means the positively charged nucleus is growing with addition negatively scattered points.

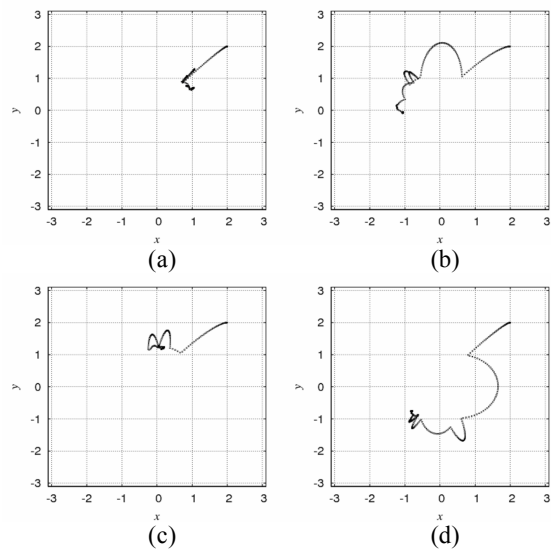

Figure 5. Trajectory of smaller particle with number: (a) 25, (b) 50, (c) 75, and (d) 100.

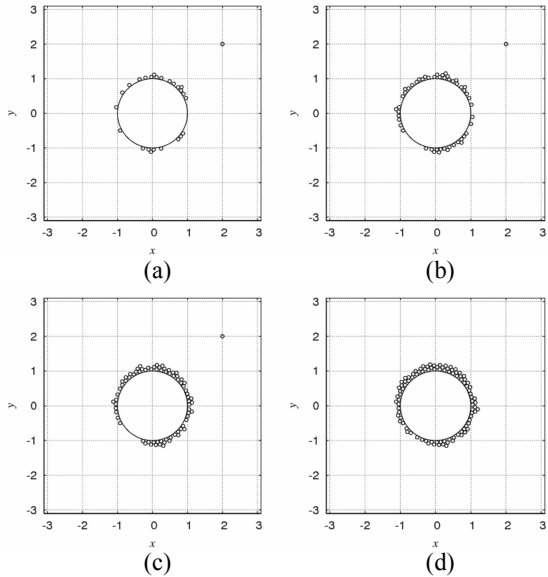

Figure 6. Deposited smaller grains on nucleus grain as seen by smaller particle with number: (a) 25, (b) 50, (c) 75, and (d) 100.

As a smaller grain injected to the system, it will see the nucleus and other previously attached smaller grains on nucleus. These circumstances will affect where it will be landed on the nucleus and then attached permanently on the nucleus. Figure 5 illustrates the trajectory of particle number 25, 50, 75, and 100, as examples, for sequence 001 (random seed 1), which is shown in the most top row in Figure 3. The related illustration about where the smaller grains attached to the nucleus is given in Figure 6. As it can be seen, the smaller grains are not deposited uniformly on the nucleus, the farthest side from injected point has less dense attached smaller grains.

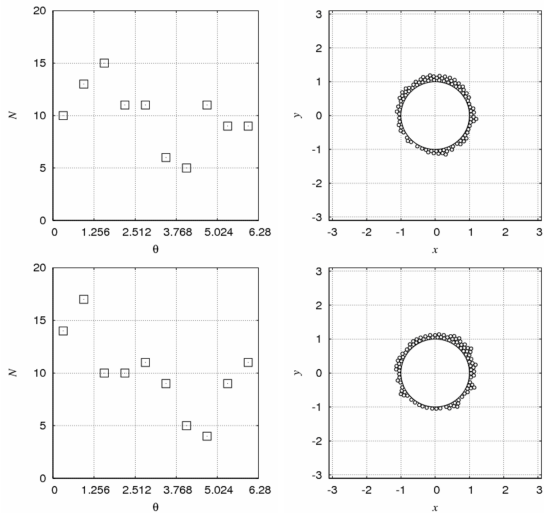

Figure 7. Angular distribution of 100 attached grains with injected position is about $\pi/4$ or 0.785 rad for: sequence 001 (top) and sequence 002 (bottom).

Further observation confirms similar result for sequence 002 (random seed 2), which is shown in the second row in Figure 3. Both results obtained from sequence 001 and 002 are given in Figure 7. It can be seen that the densest location is near the position of injected point where smaller grains enter the system but not at that point exactly since smaller grains have their initial velocity. And of course the less dense position is about $\pi/4 + \pi/2$, the other side of densest position. We observe that the sequence have influence but minor in producing final angular distribution of attached grains position of. The results should be average of all possible sequences, which is unfortunately very large [8].

## Conclusion

Deposition of smaller grains on a larger grain has been simulated. Densest site is found to be near the injected point for normal distribution of velocity with initial velocity in range of 0 and 0.02, where 15-17 % number of particles is deposited in this site.

## Acknowledgements

Author would like to thanks ITB Alumni Association research grant for supporting calculation part of this work and ITB Research Division research grant for supporting this presentation in the conference.

## References

[1] Yoji Nakajima and Takashi Nato, "Estimation of maximum charge sustainable for a spherical particle in normal air", Journal of the Institute of Electrostatics Japan 23 (2), 81-87 (1999)
[2] L. B. Schein, "Recent advances in our understanding of toner charging", Journal of Electrotatics 46 (1), 29-36 (1999)
[3] W. Stanley Czarnecki and L. B. Schein, "Electrostatic force acting on a spherically symmetric charge distribution in contact with a conductive plane", Journal of Electrostatics 61 (2), 107-115 (2004)
[4] J. -C. Laurentie, P. Traoré, C. Dragan, and L. Dascalescu, "Numerical modeling of triboelectric charging of granular materials in vibrated beds", Proceeding of Industry Application Society Annual Meeting (IAS), 2010 IEEE, Houston, Texas, USA, 3-7 October 2010, pp. 1-6
[5] S. Matsusaka, H. Maruyama, T. Matsuyama, and M. Ghadiri, "Triboelectric charging of powder: A review", Chemical Engineering Science 65 (22), 5781-5807 (2010)
[6] M. P. Allen and D. J. Tildesley, "Computer simulation of liquids", Oxford University Press, New York, 1989, pp. 82-85
[7] J. Schäfer, S. Dippel, and D. E. Wolf, "Force schemes in simulation of granular materials", Journal de Physique I 6 (1), 5-20 (1996)
[8] Veinardi Suendo and Sparisoma Viridi, "Algorithms to Produce Dicreet Gaussian Sequences", Jurnal Pengajaran Fisika Sekolah Menengah 2 (2), 19-23 (2010)
[9] Sparisoma Viridi and Veinardi Suendo, "How to produce discreet Gaussian sequences: Algorithm and code", arXiv:1107.3291v1